\documentclass[12pt]{iopart}

\usepackage{graphicx}
\usepackage{bm}

\newcommand{\beq}{\begin{equation}}
\newcommand{\eeq}{\end{equation}}
\newcommand{\beqa}{\begin{eqnarray}}
\newcommand{\eeqa}{\end{eqnarray}}

\newcommand{\ket}[1]{\left| #1 \right\rangle}
\newcommand{\bra}[1]{\left\langle #1 \right|}

\begin{document}

\title{Global Optical Control of a Quantum Spin Chain}
\author{B.~W.~Lovett}

\address{Department of Materials, Oxford University, Oxford OX1 3PH, United Kingdom}
\ead{brendon.lovett@materials.oxford.ac.uk}
\begin{abstract}
Quantum processors that combine the long decoherence times of spin qubits together with fast optical manipulation of excitons have recently been the subject of several proposals. I show here that arbitrary single- and entangling two-qubit gates can be performed in a chain of perpetually coupled spin qubits solely by using laser pulses to excite higher lying states. It is also demonstrated that universal quantum computing is possible even if these pulses are applied {\it globally} to a chain; by employing a repeating pattern of four distinct qubit units the need for individual qubit addressing is removed. Some current experimental qubit systems would lend themselves to implementing this idea.
\end{abstract}

\pacs{03.67.-a, 03.67.Lx, 78.67.Hc}

\maketitle

\section{Introduction}
Universal quantum computing (QC)~\cite{nielsen00} generally requires both the manipulation of single qubits and control of inter-qubit interactions. An enormous range of potential quantum computing hardware has been proposed, especially in solid-state systems. Many of these schemes suffer from two principle disadvantages. First, it is difficult to maintain quantum coherence in the solid phase for a time sufficiently long that enough gate operations can be performed for quantum error correction to be feasible. Second, their required addressing of individual qubits is extremely difficult. Qubits with interactions strong enough to support QC are usually only a few nanometers apart, so that connecting each one separately to a macroscopic manipulation apparatus is exceeding challenging -- and anyway introduces channels of decoherence that can destroy quantum information.

In this paper, I shall propose a way of overcoming both of these difficulties in a chain of spin qubits, each of which have an associated higher energy level that can be addressed by using a laser. The electron spin is a good choice for a qubit since it usually has a much longer coherence time than other electronic degrees of freedom in a solid~\cite{elzerman04}; many different ways of using it as the building block  of a quantum processor have been proposed~\cite{loss98, imamoglu99}. However, direct manipulation of such a qubit is a rather slow process, and so it has recently been suggested that the fast optical control of excitons in confined quantum systems be used in combination with such a quantum spin memory~\cite{pazy03, calarco03, nazir04}. In order to use the natural interactions between spins to create an entangling spin gate, it is advantageous to maximize the magnitude of their coupling by placing the spins very close to one another. For example, a chain of $n$-doped quantum dots can be fabricated with a single spin-1/2 electron in each~\cite{cortez02} and these might only be a few nanometers apart. Thus, modifying the spin-spin interaction by external means is very difficult. One way around this is to devise ways of working with spins that are continuously interacting; such a scheme has been proposed by Benjamin and Bose~\cite{benjamin03}. It is based on the use of a `barrier qubit', placed between each computational qubit, whose Zeeman splitting can be tuned by means of, say, an external magnetic field. This scheme was recently modified so that an optical pulse could be used, in conjunction with a higher lying optically active state of the barrier, to modulate the effective Zeeman energy~\cite{lovett04}. In this scheme, each barrier must be addressed individually by a laser, so making it difficult to scale.  
I shall here propose a solution to this problem, by describing a quantum spin chain that supports {\it global control}. 

\section{The Proposed Architecture}
\label{architecture}

\begin{figure}
\centering
\vspace{0cm}
\includegraphics[width=3.5in,height=2.5in]{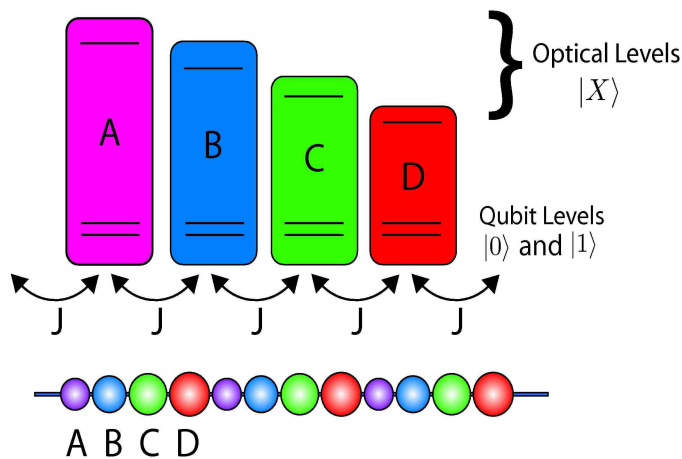}
\caption{Schematic diagram of the proposed quantum register. It consists of a repeating pattern of four distinct units. Each of these four units has an identical low energy spin qubit, together with a distinguishing higher lying level that is separated from the qubit by an optical energy. The spin qubits are perpetually coupled by an $XY$ interaction with strength $J$. $A$ and $C$ type units comprise computational qubits, whereas $B$ and $D$ units are barriers to information transfer.}
\label{scheme}
\end{figure}

Benjamin~\cite{benjamin02} showed that a device with two types of qubit, $A$ and $C$ say, can support global control so long as certain conditions are met. First, they must be arranged in the repeating sequence $ACACAC...$. Second, it must be possible to perform any unitary operation on all of the $A$ (or all of the $C$) type qubits simultaneously (without altering the state of the other qubit type). Third, alternating qubit-qubit interactions $H_{AC}$ and $H_{CA}$ must be switchable, so that an entangling operation is possible between any set of adjacent qubit pairs.

Fig.~\ref{scheme} shows an idealized version of our proposed architecture; we shall discuss real implementations towards the end of the paper. The device consists of four different units; each unit consists of two lower lying electronic spin levels (labelled $\ket{0}$ and $\ket{1}$) and a level ($\ket{X}$) at an energy corresponding to an optical frequency above them. This transition energy is different for each of the four units. Every other unit (of type $A$ and $C$ in Fig.~\ref{scheme}, say) is a computational qubit upon which algorithms can be executed, and the others are barriers which control the flow of information along the chain.

\subsection{The Building Blocks}

Let us first consider a short three-unit $ABC$ section of our device; we shall discuss scaling up subsequently. The Hamiltonian of this three unit section is written ($\hbar=1$):
\beqa
{\cal H} &=& \sum_{i \in \{A, B, C\}} \left(\omega_0 \ket{1_i}\bra{1_i}+ \omega_1^i\ket{X_i}\bra{X_i}\right)\\
&&+ J\left(\ket{0_A1_B}\bra{1_B0_A} + \ket{0_B1_C}\bra{1_B0_C} +H.c.\right)\nonumber
\eeqa
where $\omega_0$ is the Zeeman splitting of the spin qubits, $\omega_1^i$ is the optical transition energy for each unit $i$ and $J$ is the $XY$ exchange coupling between adjacent spins. 

We first consider the required entangling operation between $A$ and $C$. In Ref.~\cite{lovett04}, we demonstrated that a laser applied such that it resonantly couples only the $\ket{0}$ and $\ket{X}$ levels of the barrier ($B$) will take the barrier spin out of resonance with qubits $A$ and $C$. This prevents the interaction $J$ from causing energy to transfer between the three spins 
(so long as the Rabi frequency describing the coupling strength of the laser is significantly larger than $J$) -- and it therefore passivates the device. Let us assume that the barrier is initialized in state $\ket{1}$. Switching the laser off for a time $t_R = 2\sqrt{2}\pi/J$ returns the barrier to state $\ket{1}$ and effects the entangling operation:
\beq
\label{2qgate}
U_e=\left(
\begin{array}{cccc}
1 &0 &0 &0\\
0 &0 &-1 &0\\
0 &-1 &0 &0\\
0 &0 &0 &-1
\end{array}
\right).
\eeq
between the two qubits $A$ and $C$~\cite{benjamin03}.

Whenever a single qubit operation is to be performed, we must prevent energy transfer between the qubits by applying a laser to the barrier. We then reduce the problem to one of isolated, decoupled qubits. Arbitrary single qubit manipulation requires the ability to perform a rotation of any angle about two different axes of the Bloch sphere~\cite{nielsen00}. There are several ways of achieving this in our system; let us focus here on an $x$- and $z$-axis rotation. These can be written as $R_i(\theta) = \exp(i\sigma_i\theta/2)$; $\sigma$ denotes a Pauli matrix and $i\in\{ x, z\}$. 

The $x$-rotation may be achieved by performing a Raman transition between the two lower (qubit) levels; two lasers are used, each of which addresses a transition between one of the qubit levels and the higher state. 
In our idealized model, let us assume that it is possible to exploit angular momentum selection rules to couple one laser to the $\ket{0}-\ket{X}$ transition and the other to $\ket{1}-\ket{X}$. The Hamiltonian of qubit $A$ (which is decoupled from the others by the activated barrier) is:
\beqa
{\cal H}_A &=& \omega_0 \ket{1}\ket{1} + \omega_1^A \ket{X}\bra{X} + \Omega_1 \cos(\omega_{l1} t ) (\ket{0}\bra{X} + H. c)\nonumber\\ &&+ \Omega_2 \cos(\omega_{l2} t ) (\ket{1}\bra{X} + H. c)
\eeqa
$\Omega_{1(2)}$ is the laser coupling strength for the $\ket{0}-\ket{X}$ ($\ket{1}-\ket{X}$) transition. $\omega_{l1(l2)}$ is the laser frequency for the laser addressing the
$\ket{0}-\ket{X}$ ($\ket{1}-\ket{X}$) transition. Each laser is detuned from its respective transition by an amount $\delta$ (i.e. $\delta = \omega_{l1} - \omega_1 = \omega_{l2} -\omega_1 + \omega_0$). In this case, we can move to a rotating frame and make the rotating wave approximation, and the Hamiltonian can be rewritten:
\beq
{\cal H}_A = \delta \ket{X}\bra{X} + \frac{\Omega_1}{2} (\ket{0}\bra{X} + H. c) + \frac{\Omega_2}{2} (\ket{1}\bra{X} + H. c).
\eeq
Assuming that $\delta \gg \Omega_1, \Omega_2$, a straightforward degenerate perturbation theory calculation reveals an effective coupling between $\ket{0}$ and $\ket{1}$ of strength $\Omega_1 \Omega_2/2\delta$.
Applying the pulses for a time $\tau$ therefore results in the operation $R_x (\Omega_1 \Omega_2\tau/2\delta)$, and by varying $\tau$ any rotation angle is possible.

The $z$-rotation can be achieved by using a single laser which couples the system between one of the qubit levels and the higher state, which might again be achieved by exploiting angular momentum selection rules. In this case, we only use one laser and so our Hamiltonian may be written:
\beq
{\cal H}_A = \omega_0 \ket{1}\bra{1} + \omega_1^A \ket{X}\bra{X} + \Omega_1\cos(\omega_{l1} t +\phi_1) (\ket{0}\bra{X} + H. c)
\eeq
$\phi_1$ describes the laser phase. $\omega_{l1}$, the laser frequency, is set equal to $\omega_1^A$, the $\ket{0}-\ket{X}$ transition energy. After moving into a frame rotating at this frequency and making the rotating wave approximation, we obtain:
\beq
{\cal H}_A = \omega_0 \ket{1}\bra{1} + \frac{\Omega_1}{2} (\ket{0}\bra{X}\exp(-i\phi_1) + H. c)
\eeq
Now consider a specific situation: that we are initially in the qubit state $\ket{0}$ with the laser turned off. We now apply a pulse for a time $\tau = \pi/\Omega$, and all the population moves to $\ket{X}$ (this is a $\pi$ pulse). A second $\pi$ pulse is now applied, this time with phase parameter $\phi_2$. This results in a transformation, in the lab frame, of $\ket{0}\rightarrow -\exp(i(\phi_2-\phi_1))\ket{0}$. The natural evolution of $\ket{1}$ is unaffected by the pulses since it does not couple to them. Therefore, an arbitrary superposition of $\ket{0}$ and $\ket{1}$ undergoes a $z$-rotation on the Bloch sphere whose phase angle is determined by the phase difference between the two $\pi$ laser pulses: $R_z(\phi_2-\phi_1+\pi)$.

Since each of our three units has a different optical transition we can use frequency selectivity to address each one separately. This allows us to perform single qubit gates separately on both qubits $A$ and $C$ as well as controlling their interaction by manipulating the state of $B$. (Though we have neglected the question of how a pulse applied to qubit $A$ might affect qubit $C$; we shall return to this in Sec.~\ref{implementation}.) This set of gates is universal for our two qubits.

\subsection{An Example: The CNOT gate}

\begin{figure}
\centering
\vspace{0cm}
\includegraphics[width=3in,height=1.2in]{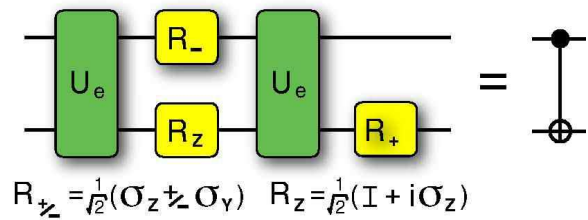}
\caption{Pictorial representation of the sequence required to produce a CNOT gate from the entangling operation (Eq.~\ref{2qgate}) and single qubit operations.}
\label{CNOTfig}
\end{figure}

\begin{figure}
\centering
\vspace{0cm}
\includegraphics[width=3in,height=5.5in]{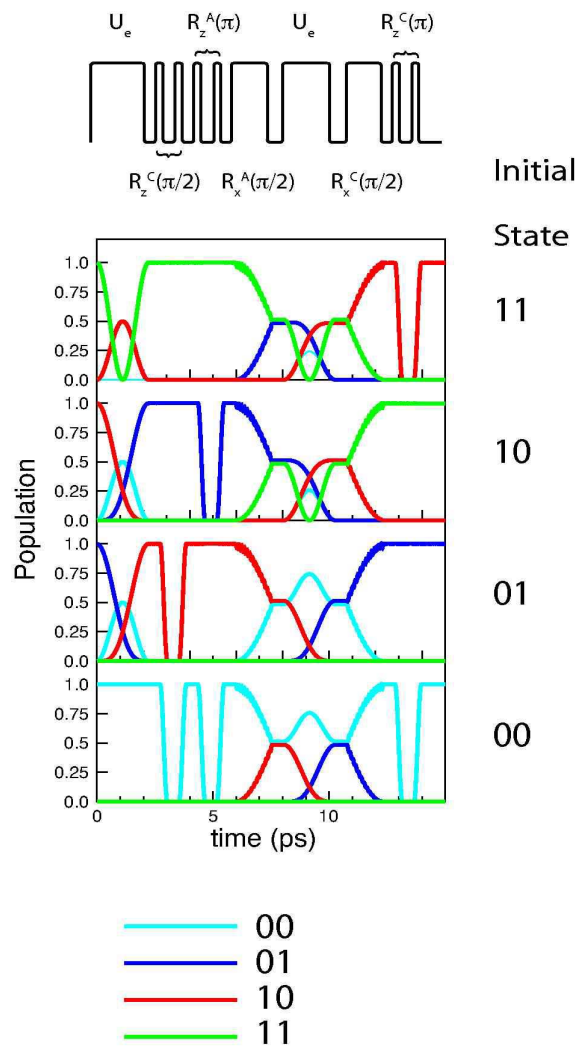}
\caption{Top: pulse sequence required to implement a CNOT gate using the basic operations discussed in the text. Bottom: CNOT gate dynamics for an initial state of 00, 01, 10, 11. The parameters are as follows. $J = 1$~THz; for the Raman $X$ gate: $\Omega_1 = \Omega_2 =  10$~THz, $\nu = $ 50~THz; for the selective excitation two pulse $Z$ gate: $\Omega_1 = 10$~THz; for the $J$ passivation $\Omega= 100$~THz. The procedure for applying each pulse and for how long each pulse should be applied is described in the text.}
\label{CNOT}
\end{figure}

In order to demonstrate that all of these gates can work together, let us consider the action of the following sequence of gate operations which make up a CNOT gate (see Fig.~\ref{CNOTfig} for a pictorial representation of this sequence):
\beq
U_{CNOT} = R^C_z(\pi)R_x^C\left(\frac{\pi}{2}\right)U_eR_x^A\left(\frac{\pi}{2}\right)R^A_z(\pi)R^C_z\left(\frac{\pi}{2}\right)U_e.
\eeq
Fig.~\ref{CNOT} shows the populations of the four computational basis states as a function of time during the pulse sequence, after the register has been initialized to each of these same states. The CNOT gate is executed with high fidelity. 

\subsection{Scaling Up: Global Control}

Let us now reconsider the complete repeating structure of Fig.~\ref{scheme}. Applying the laser to all of the barriers $B$ blocks all the interactions $H_{AC}$. Similarly, the $H_{CA}$ are blocked by addressing all of the $D$ type barriers. When one of these sets of interactions is blocked, the chain is divided into groups of three units (two qubits and a barrier), on which, as we have just demonstrated, any two qubit evolution can be executed. We have therefore satisfied the requirements laid out earlier and therefore conclude that our {\it four-unit qubit-barrier structure supports global control.}

\section{Decoherence}

Decoherence is a key issue for any quantum computing scheme. The shortest decoherence time in our system is that for spontaneous emission of photons from the uppermost (optical) level, and we must try and minimise the effect of this. Let us examine how each gate operation is affected by this kind of decoherence.

The operation $U_e$ does not involve the higher optical levels at all, but when the entangling interaction must be passivated, emission of photons is possible. However, in Ref.~\cite{lovett04}, we showed that during passive periods, the device remains very robust to effect of this process. The fidelity of the qubits remains greater than $0.995$ for a time equal to the entangling gate period, when the optical decay time is of the same order of magnitude as the gate period.

It is straightforward to increase the decoherence time of the $x$ rotation that is controlled by a Raman process~\cite{nazir04}. In order to show this, let us consider the case where $\Omega_1 = \Omega_2 = \Omega$. By simply making the ratio $\alpha \equiv \Omega/\delta$ smaller, the population of $\ket{X}$ excited during the Raman process is reduced since it is $\leq 4\alpha^2$. If the natural decoherence time of the $\ket{X}$ state is $\tau_X$ (which might typically be a nanosecond~\cite{borri01}), then a crude estimate of the system decoherence time is now $\tau_d = \tau_X/4\alpha^2$. The drawback of this method is that the gate operation now takes longer. For example, a $R_x(\pi)$ operation takes $\tau_g = \pi/\Omega\alpha$. However, the figure of merit for quantum computing is the ratio of these two times, and this is increased as $\Omega\tau_X/4\pi\alpha$. 

\begin{figure}
\centering
\vspace{0cm}
\includegraphics[width=3in,height=3in]{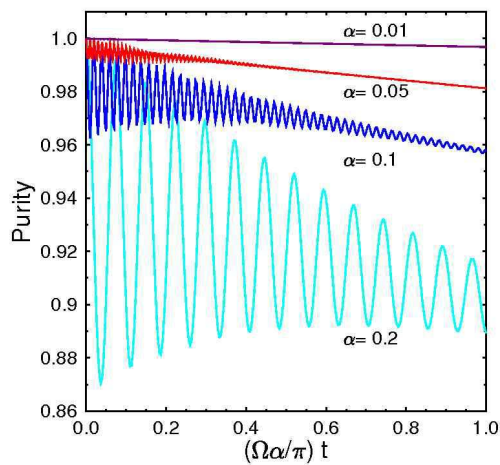}
\caption{Purity of the qubit density matrix as a function of time during a $R_z(\pi)$ gate. I take $\tau_X = 0.1/\Omega$ and show plots for various $\alpha = \Omega/\delta$ (see text for definitions).}
\label{Zgate}
\end{figure}

We adopt a similar strategy for improving the coherence characteristics the $z$ rotation operation, by using a more complicated protocol for performing this gate than that 
described earlier. Let us again consider a transition between one of the qubit levels, $\ket{0}$ say, and $\ket{X}$. If a laser with coupling strength $\Omega$ is tuned to this resonance, and a $2\pi$ pulse applied to an initial $\ket{0}$, a phase of $\pi$ is acquired. If this laser is now detuned by an amount $\delta$ such that, as before, $\alpha = \Omega/\delta \ll 1$, then after a time $2\pi/\alpha$ the population returns to $\ket{0}$ but the phase picked up is now $2\pi\alpha^2$. A $R_z(\pi)$ gate is possible provided that this detuned pulse is repeated $1/2\alpha^2$ times; this takes a total time $\tau_g = \pi/\delta\alpha^2$, and the figure of merit is then $\tau_X\delta/4\pi$. Thus, we again find that {\it increasing the detuning increases the figure of merit}. This effect is displayed in Fig.~\ref{Zgate}; the purity is defined as the trace of the square of the reduced density matrix for the qubit. I show this for an initial state of $2^{-1/2}(\ket{0} + \ket{1})$ as a function of time during the described $R_z(\pi)$ gate. I assume that the decay of the optical state can be modelled by using a Markovian master equation~\cite{nielsen00}, and that $\tau_X = 0.1/\Omega$.  The improved purity for smaller $\alpha$ is a direct consequence of the dependence on $\alpha$ of the figure of merit.

The other requirements for a scalable quantum computer, initialization and measurement, may also be achieved by using the resonant fluorescence of the spin-dependent optical transition~\cite{nazir04}.

\section{Implementation in Real Systems}
\label{implementation}

The repeating four unit structure required for this scheme could be achieved in a variety of different nanoscopic systems. For example, self-assembled quantum dots can be doped with an electron spin~\cite{cortez02}. The optical creation of trions depends on the spin state if circularly polarized light is used~\cite{lovett05, imamoglu05}, and this facilitates the entangling gate we have discussed. However, the angular momentum selection rules in this system do {\it not} allow for two lasers to be applied such that each couples one of the spin states to a single higher level, and so we shall here discuss a different way of performing single qubit Raman ($x$-rotation) gates optically in these materials. We shall also show that the same structure can support $z$-rotations.

\subsection{Real Single Qubit Gates}
Let us consider the higher optical levels consisting of light hole trion states (i.e. states composed of two electrons and one light hole). The wavefunction of both types of particle in a nanostructure may be represented by:
\beq
\psi = U \phi
\eeq
where $U$ is a function with the periodicity of the underlying lattice, and $\phi$ is an envelope function, which describes the modulation of the wavefunction due to the potential imposed by the nanostructure. Henceforth, we shall only consider envelopes with $s$ symmetry, which give rise to states with the lowest energies. We can represent the $U$ states of electrons and light holes as follows:
\beqa
\label{bloch1}
\ket{3/2_h, 1/2} &=& \frac{f(r)}{\sqrt{6}}\left[\ket{(X+iY)\beta} 
- \ket{2Z\alpha}\right],\\
\ket{3/2_h, -1/2} &=& \frac{f(r)}{\sqrt{6}}\left[\ket{(X-iY)\alpha} 
+ \ket{2Z\beta}\right],\\
\ket{1/2_e, 1/2} &=&g(r)\ket{S\alpha},\\
\ket{1/2_e, -1/2} &=&g(r)\ket{S\beta}.
\eeqa
We have labeled the Bloch functions $U$ by using the notation 
$\ket{J_p, J_z}$ for particle type $p$, total angular momentum $J$ and $z$ angular momentum projection $J_z$.  
$f$ and $g$ describe radial dependence;
 $\alpha$ and $\beta$ are the up and down spin states respectively. 
The $X$, $Y$, $Z$ and $S$ represent orbital wavefunctions as follows:
\beqa
\bra{\bf r}X\rangle&=& \sqrt{\frac{3}{4\pi}}\sin\theta\cos\phi\\
\bra{\bf r}Y\rangle&=& \sqrt{\frac{3}{4\pi}}\sin\theta\sin\phi\\
\bra{\bf r}Z\rangle&=& \sqrt{\frac{3}{4\pi}}\cos\theta\\
\bra{\bf r}S\rangle&=& \sqrt{\frac{1}{4\pi}}
\eeqa

We want to create excitons composed of electrons and light holes by using a laser, and the relevant coupling is through the dipole operator. Consider first creating excitons from the vacuum state. In this case, we obtain the following dipole matrix elements for the various possible transitions:
\beqa
\ket{vac} \rightarrow \ket{\frac{1}{2}_h, -\frac{1}{2}_e}&:& M_{\bf p} = \frac{2A}{\sqrt 6}{ \bf k}\label{trans1}\\
\ket{vac} \rightarrow \ket{-\frac{1}{2}_h, \frac{1}{2}_e}&:& M_{\bf p} = -\frac{2A}{\sqrt 6}{ \bf k}\\
\ket{vac} \rightarrow \ket{\frac{1}{2}_h, \frac{1}{2}_e}&:& M_{\bf p} = \frac{A}{\sqrt 6}({\bf i} - i{\bf j})\label{trans3}\\
\ket{vac} \rightarrow \ket{-\frac{1}{2}_h, -\frac{1}{2}_e}&:& M_{\bf p} =\frac{A}{\sqrt 6}({\bf i} + i{\bf j})
\label{trans}
\eeqa
$A$ is a constant for a specific dot.
If a laser is applied with polarization in the $yz$ plane, its field can be represented as follows:
\beq
{\bf E} = (\sin(\theta){\bf j}+ \cos(\theta){\bf k})E_0 \cos(\omega_l t)
\eeq
$\omega_l$ is the laser frequency, $E_0$ the amplitude of the laser field, and $\theta$ describes the orientation of the polarization in the $yz$ plane.
The field interacts with the dipole in the usual way. 

Let us consider what happens when the dot is doped with a single extra electron spin. In this case it is only possible to create a light hole exciton when the created electron spin is oppositely directed to the spin that already exists in the dot. We may then write our Hamiltonian in a basis which includes the two possible electron spin states, and the two trion states that can be excited by our laser. After moving into a frame rotating at $\omega_l$ and making the rotating wave approximation  the Hamiltonian reads
\beq
\label{spinham}
{\cal H}= \left( \begin{array}{cccc}
0 & 0 & \Omega\cos\theta & 2\Omega\sin\theta \\
0 & 0 & -2\Omega\sin\theta & -\Omega\cos\theta \\  
\Omega\cos\theta & -2\Omega\sin\theta & \delta & 0\\   
2\Omega\sin\theta & -\Omega\cos\theta & 0 & \delta \\  \end{array} \right).
\eeq
If we now switch to a notation that simply depicts the $z$ projection of the electrons or holes as an arrow, the states are in the order $\ket{\uparrow_e}, \ket{\downarrow_e}, \ket{\uparrow_e, \downarrow_e, \uparrow_h},  \ket{\uparrow_e, \downarrow_e, \downarrow_h}$. The laser-dot coupling $\Omega = E_0 A/\sqrt{6}$.

If we assume that $|\Omega\sin\theta |, |\Omega\cos\theta | \ll \delta$, we can
apply degenerate perturbation theory to the qubit subspace to obtain an effective Hamiltonian in this subspace. It reads:
\beq
\label{spinham}
{\cal H}_{eff}=\frac{\Omega^2}{\delta} \left( \begin{array}{cc}
1+\sin^2\theta & \sin 2\theta\\
\sin 2\theta & 1+\sin^2\theta  
\end{array} \right).
\eeq
The diagonal terms are the AC Stark effect shifts, and may be ignored since they are the same for each qubit state. The off-diagonal terms represent and effective coupling between $\ket{\uparrow_e}$ and $\ket{\downarrow_e}$ and are due to the Raman effect, which in this case goes via two higher states. By pulsing the laser, we modulate $\Omega$ and can therefore perform a rotation of any angle about the $x$-axis of the Bloch sphere.

As discussed earlier, for universal quantum computing we also require another rotation on the Bloch sphere. This can be done with circularly polarized light, whose field can be represented by:
\beq
{\bf E} = \Re \left(\frac{{\bf i}+i{\bf j}}{\sqrt{2}}E_0 \exp(i\omega_l t)\right).
\eeq
The only non-zero dipole matrix element for this kind of field, of those listed in Eqs.~\ref{trans1} to~\ref{trans} is that of Eq.~\ref{trans3}. With an extra electron in the dot representing our qubit, it is therefore possible to have a spin selective transition; an exciton is only created from the spin down state (and further, only one type of exciton is created). It is possible therefore to use exactly the method outlined in Section~\ref{architecture} to rotate around the $z$-axis of the Bloch sphere.

The lifetime of an electron spin in a doped self-assembled dot system has been measured to be over 10~ms~\cite{kroutvar04}, whereas excitonic lifetimes are on the nanosecond scale~\cite{borri01}; it would be sensible therefore to employ the detuning techniques discussed earlier to improve the decoherence characteristics of the device. A direct spin-spin interaction of about 1~meV has been seen in a lithographically defined structure~\cite{chen04}, which would allow an entangling gate to be performed in a few picoseconds. Self-organized quantum dots can be grown in regularly spaced stacks~\cite{cortez02} and the excitonic transition energy varies naturally as the stack size increases~\cite{tersoff96}. The energy can also be controlled by varying the growth conditions in a typical molecular beam epitaxy set-up.  
Apparently then, we are able to perform arbitrary single qubit gates in this system. However, in the next part we shall discuss a potential problem, and how to resolve it.

\subsection{Addressing Two Qubits Simultaneously} 

If we are using a laser to perform single qubit operations on the computational qubits of our quantum spin chain, we must be able to manipulate only qubits of type $A$ or qubits of type $C$ individually. However, each laser will impinge on both types of qubits, since it has much longer wavelength than the interqubit separation. For example, consider a Raman type pulse acting on qubit $A$, with a detuning $\delta_A$. From Eq.~\ref{spinham} it induces Rabi oscillations in the qubit of frequency:
\beq
f_A = \frac{\Omega^2}{\delta_A} (\sin 2\theta).
\eeq

Assuming that the coupling strength $\Omega$ and the polarization angle $\theta$ does not very from dot to dot, the frequency of Rabi oscillations induced in qubit $C$ is:
\beq
f_C = \frac{\Omega^2}{\delta_C} (\sin 2\theta).
\eeq
A pulse resulting in a phase angle $\phi_A$ on qubit $A$ will therefore result in a simultaneous operation on qubit C, of phase angle
\beq
\phi_C = \frac{\delta_A}{\delta_C}\phi_A.
\eeq

\begin{figure}[h]
\centering
\vspace{1cm}
\includegraphics[width=3in,height=4in]{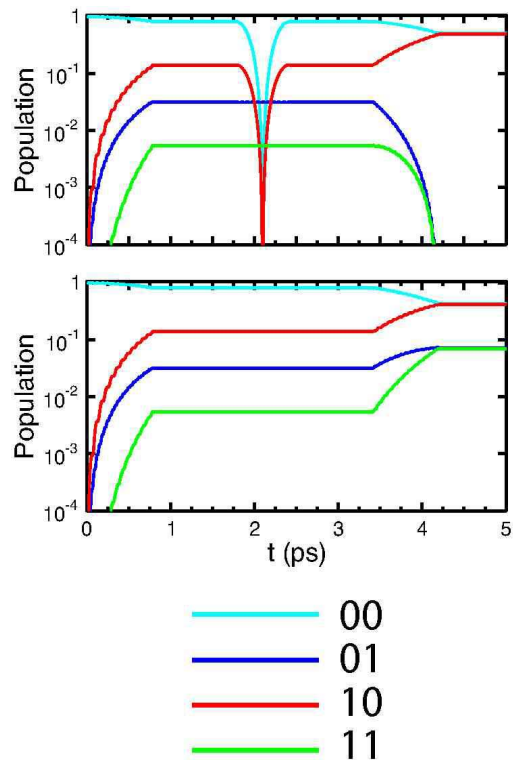}
\caption{Population of the computational basis states for two qubits manipulated through their optical transitions. The lower figure shows the effect of an uncorrected $R_x^A(\pi/2)$ pulse on an initial state $\ket{00}$. The upper figure shows the effect of an $R_x^A(\pi/2)$ on $\ket{00}$, corrected using the method described in the text. Parameters are $\Omega = 5$~THz, $\theta = 26^\circ$, $\delta_A = 100$~THz, and $\delta_C = 100$~THz. A laser-dot coupling $\Omega^\prime=5$~THz is used to effect the correcting $z$ rotation.}
\label{singlecorrect}
\end{figure}

This is clearly a problem. One solution would be to make the ratio $\delta_A/\delta_C$ very small (i.e. detune $C$ by a much greater amount than $A$). This may not always be possible however, since a highly detuned laser may interfere with other levels of the quantum dot. We therefore accept this error, but correct it. Instead of performing the rotation $R_x(\phi_A)$ in one step, we split it into two and insert an $R_z(\pi)$ pulse, on qubit $C$. Since our method for applying rotations can be done resonantly, the $R_z(\pi)$ pulse has a negligible effect on qubit $A$, and means that any erroneous phase built up in the $z$ rotation on qubit $C$ is cancelled. A final $R_z(\pi)$ pulse on qubit $C$ returns it to its starting position. We show this effect in Fig.~\ref{singlecorrect}.

\section{Molecular Embodiments}

Though quantum dots form an ideal system in which to test the scheme I have presented here, molecules offer a more promising long term route towards a large quantum computer. These offer both the potential for stronger interqubit interactions (courtesy of the shorter interqubit distances that may be achieved), and the possibility of arrays of identical repeating patterns. For example, spin-active endohedral fullerenes can be arranged in a one dimensional pattern inside a carbon nanotube~\cite{khlobystov04}. N@C$_{60}$ has a spin dephasing time of at least 240~$\mu$s~\cite{morton05}, and other spin active species have been shown to exhibit magneto-optical activity~\cite{jones05}. Spin resonance measurements indicate that the spin-spin interactions between fullerenes could be large enough for a two qubit gate to be performed on the sub-$\mu$s timescale~\cite{jakes03}.
Many different atoms and molecules have been put inside fullerene cages~\cite{shinohara00}, and these have a range of different optical properties. It is therefore possible that a four unit structure, such as that proposed here, could be synthesized.

\section{Summary}

To summarize, I have demonstrated a new scheme for global optical control of a quantum spin chain, which supports universal quantum computing. 
The paper has shown how to combine the long spin decoherence time, the fast optical manipulation of excitons and the convenience of global control (which leads to stronger qubit-qubit interactions) into a single device. I hope that these features will drive experimental progress towards implementing the scheme in quantum dots or molecular systems.

\ack
I would like to thank S.~C.~Benjamin, G.~A.~D.~Briggs, T.~P.~Spiller and S.~D.~Barrett for stimulating discussions. This research is part of the QIP IRC www.qipirc.org (GR/S82176/01) and is supported through DSTL.


\section*{References}
\begin{thebibliography}{23}
\expandafter\ifx\csname natexlab\endcsname\relax\def\natexlab#1{#1}\fi
\expandafter\ifx\csname bibnamefont\endcsname\relax
  \def\bibnamefont#1{#1}\fi
\expandafter\ifx\csname bibfnamefont\endcsname\relax
  \def\bibfnamefont#1{#1}\fi
\expandafter\ifx\csname citenamefont\endcsname\relax
  \def\citenamefont#1{#1}\fi
\expandafter\ifx\csname url\endcsname\relax
  \def\url#1{\texttt{#1}}\fi
\expandafter\ifx\csname urlprefix\endcsname\relax\def\urlprefix{URL }\fi
\providecommand{\bibinfo}[2]{#2}
\providecommand{\eprint}[2][]{\url{#2}}

\bibitem{nielsen00}
Nielsen M A and Chuang I L 2000 {\it Quantum Computation and Quantum Information} (Cambridge: Cambridge University Press)

\bibitem{elzerman04}
Elzerman J M, Hanson R, van Beveren L H W, Witkamp B, Vandersypen L M K and Kouwenhoven L P 2004 {\it Nature} {\bf 430} 431

\bibitem{loss98}
Loss D and DiVincenzo D P 1998 {\it \PR} A {\bf 57} 120


\bibitem{imamoglu99}
Imamoglu A, Awschalom D D, Burkard G, DiVincenzo D P, Loss D, Sherwin M and Small A 1999 {\it \PRL} {\bf 83} 4204

\bibitem{pazy03}
Pazy E, Biolatti E, Calarco T, D'Amico I, Zanardi P, Rossi F and Zoller P 2003 {\it Europhys.~Lett.} {\bf 62} 175

\bibitem{calarco03}
Calarco T, Datta A, Fedichev P, Pazy E and Zoller P 2003 {\it \PR} A {\bf 68} 012310

\bibitem{nazir04}
Nazir A, Lovett B W, Barrett S D, Spiller T P and Briggs G A D 2004 {\it \PRL} {\bf 93} 150502

\bibitem{cortez02}
Cortez S, Krebs O, Laurent S, Senes M, Marie X, Voisin P, Ferriera R, Bastard G, G\'erard J-M and Amand T 2002 {\it \PRL} {\bf 89} 207401

\bibitem{benjamin03}
Benjamin S C and Bose S 2003 {\it \PRL} {\bf 90} 247901

\bibitem{lovett04}
Benjamin S C, Lovett B W and Reina J H 2004 {\it \PR} A {\bf 70} 060305(R)

\bibitem{benjamin02}
Benjamin S C 2002 {\it \PRL} {\bf 88} 017904

\bibitem{benjamin04}
Benjamin S C and Bose S 2004 {\it \PR} A {\bf 70} 032314


\bibitem{borri01}
Borri P, Langbein W, Schneider S, Woggon U, Sellin R L, Ouyang D and Bimberg D 2001 {\PRL} {\bf 87} 157401

\bibitem{lovett05}
Lovett B W, Nazir A, Pazy E, Barrett S D, Spiller T P and Briggs G A D 2005 {\it \PR} B {\bf 72} 115324

\bibitem{imamoglu05}
Hogele A, Kroner M, Seldi S, Karrai K, Atature M, Dreiser J, Imamoglu A, Warburton R J, Badolato A, Gerardot B D and Petroff P M 2005 {\it Appl. Phys. Lett.} {\bf 86} 221905

\bibitem{kroutvar04}
Kroutvar M, Ducommun Y, Hess D, Bichler M, Schuh D, Abstreiter G and Finlay J J 2004 {\it Nature} {\bf 432} 81

\bibitem{chen04}
Chen J C, Chang A M and Melloch M R 2004 {\it \PRL} {\bf 92} 176801

\bibitem{tersoff96}
Tersoff J, Teichert C and Lagally M G 1996 {\it \PRL} {\bf 76} 1675

\bibitem{khlobystov04}
Khlobystov A N, Britz D A, Ardavan A and Briggs G A D 2004 {\it \PRL} {\bf 92} 245507

\bibitem{morton05}
Tyryshkin A M, Morton J J L, Ardavan A, Porfyrakis K, Lyon S A and Briggs G A D {\it unpublished}

\bibitem{jones05}
Jones M A G and Morton J J L {\it unpublished}

\bibitem{jakes03}
Jakes P, Dinse K P, Meyer C, Harneit W and Weidinger A 2003 {\it Phys. Chem. Chem. Phys.} {\bf 5} 4080

\bibitem{shinohara00}
Shinohara H 2000 {\it Rep. Prog. Phys.} {\bf 63} 844

\end{thebibliography}
\end{document}